\documentclass[sigconf]{acmart}

\usepackage[para]{footmisc}

\usepackage{amsmath}
\usepackage{natbib}
\usepackage{amsfonts}
\usepackage{graphicx}
\usepackage{hhline}
\usepackage{hyperref}
\usepackage{enumerate}
\usepackage{multirow}
\usepackage{array}
\usepackage[para]{footmisc}
\usepackage{adjustbox}
\usepackage{balance}
\usepackage{subcaption}
\usepackage[normalem]{ulem}
\usepackage[show]{chato-notes}
\usepackage{marginnote}
\newcommand{\pageenlarge}[1]{\enlargethispage{#1\baselineskip}}
\usepackage{subcaption}

\usepackage{algorithm}
\usepackage{algorithmicx}
\usepackage{algpseudocode}

\copyrightyear{2023}
\acmYear{2023}
\setcopyright{acmcopyright}\acmConference[WSDM '24]{32nd ACM International Conference on Information and Knowledge Management}{March 4--8, 2024}{Mérida, Mexico}
\acmBooktitle{WSDM '24), March 4--8, 2024, Mérida, Mexico}

\newcommand{\craig}[1]{\textcolor{black}{#1}}

\newcommand{\io}[1]{\textcolor{black}{#1}}
\newcommand{\yo}[1]{\textcolor{black}{#1}}

\newcommand{\zixuan}[1]{\textcolor{black}{#1}}

\newcommand{\zijun}[1]{\textcolor{black}{#1}}
\newcommand{\ric}[1]{\textcolor{black}{#1}}
\newcommand{\zxrevise}[1]{\textcolor{black}{#1}}
\newcommand{\ya}[1]{\textcolor{black}{#1}}
\newcommand{\zijunrevise}[1]{\textcolor{black}{#1}}
\newcommand{\cmw}[1]{\textcolor{black}{#1}}

\begin{document}

\title{Large Multi-modal Encoders for Recommendation}

\author{Zixuan Yi}
\email{z.yi.1@research.gla.ac.uk}
\affiliation{%
 \institution{University of Glasgow}
 \city{Glasgow}
 \state{Scotland}
 \country{United Kingdom}
}

\author{Zijun Long}
\email{z.long.2@research.gla.ac.uk}
\affiliation{%
 \institution{University of Glasgow}
 \city{Glasgow}
 \state{Scotland}
 \country{United Kingdom}
}

\author{Iadh Ounis}
\email{iadh.ounis@glasgow.ac.uk}
\affiliation{%
 \institution{University of Glasgow}
 \city{Glasgow}
 \state{Scotland}
 \country{United Kingdom}
}

\author{Craig Macdonald}
\email{craig.macdonald@glasgow.ac.uk}
\affiliation{%
 \institution{University of Glasgow}
 \city{Glasgow}
 \state{Scotland}
 \country{United Kingdom}
}

\author{Richard Mccreadie}
\email{richard.mccreadie@glasgow.ac.uk}
\affiliation{%
 \institution{University of Glasgow}
 \city{Glasgow}
 \state{Scotland}
 \country{United Kingdom}
}


\fancyhead{}
\begin{abstract}
In recent years, the rapid growth of online multimedia services, such as e-commerce \ric{platforms}, has necessitated \ric{the development of personalised recommendation approaches that can encode diverse content about each item. Indeed, modern} multi-modal recommender systems exploit diverse features obtained from raw images and item descriptions to enhance \io{the} recommendation performance. However, \io{the} existing multi-modal recommenders primarily depend on the features extracted individually \io{from different media} through pre-trained modality-specific encoders, and exhibit only shallow alignments between different modalities -- limiting these systems' \io{ability} to capture \io{the} underlying relationships between the modalities.
\zxrevise{In this paper, we investigate the usage of large multi-modal encoders within the specific context of recommender systems, as these have previously demonstrated state-of-the-art effectiveness when ranking items across various domains.}
\zxrevise{Specifically, we \ya{tailor} two state-of-the-art multi-modal encoders (CLIP and VLMo) for recommendation tasks using a range of strategies, including the exploration of pre-trained and fine-tuned encoders, as well as the assessment of the end-to-end training of these encoders.
We demonstrate that pre-trained large multi-modal encoders can generate more aligned and effective user/item representations compared to existing modality-specific encoders across three multi-modal recommendation datasets.
Furthermore, we show that fine-tuning these large multi-modal encoders with recommendation datasets leads to an enhanced recommendation performance.
In terms of different training paradigms, our experiments 
\ya{highlight} the essential role of \ya{the} end-to-end training of large multi-modal encoders in multi-modal recommendation systems.}

\end{abstract}

\maketitle

\section{Introduction}\label{sec:intro}
\looseness -1 \zixuan{Traditional recommendation systems primarily rely on user-item interactions to provide personalised recommendations, which may not fully capture the rich information embedded in the diverse forms of data associated with items, such as images, texts, and audio~\cite{zhou2023comprehensive}. 
Multi-modal recommendation systems \ya{address} this issue by \ric{representing items using \zijunrevise{encodings} from multiple modalities, and hence provide more effective recommendations}~\cite{liu2023multimodal}.
By incorporating various modalities, multi-modal recommenders~\cite{liu2022multi,pan2022multimodal,chen2019personalized,kim2022mario} bridge the gap between general recommendation systems and the complexities of multimedia item content.}
\zxrevise{However, \ya{the} existing multi-modal recommenders~\cite{liu2022multi,pan2022multimodal,kim2022mario} fuse \ya{the} extracted multi-modal features into user/item representations
without sufficiently addressing the complex and inherent correlations between different modalities~\cite{wei2023lightgt}. 
For example, 
MMGCL and LATTICE (as detailed in Table 1), fail to effectively fuse the multi-modal features. Contrary to our expectations, these models showed \ya{a suboptimal performance} when integrating \ya{the} visual and textual features compared to \ya{when} using single modality data, such as text or visual features independently.}
\zxrevise{We argue that the suboptimal performance
is caused by the shallow alignment, since the used methods cannot address the complex and inherent correlations between
these modalities.}


{{\zijunrevise{Recent progress in multi-modal learning has been concentrated on developing large multi-modal (LMM) encoder architectures. These structures are designed to enable \zxrevise{a deeper} alignment of embeddings across different modalities (refer to Fig.~1).} \zijunrevise{From a structural standpoint, such LMM encoders may be {dual}-stream, such as CLIP~\cite{radford2021learning}, or single-stream (unified), such as VLMo~\cite{bao2022vlmo} (see Fig 2), where both types aim to mitigate information loss and can capture cross-modal interactions~\cite{radford2021learning}.}  Indeed, LMM encoders have been shown to result in \ya{an} increased downstream effectiveness on tasks such as image-text retrieval~\cite{rao2022does} and zero-shot classification~\cite{bao2022vlmo}. However, to-date, LMM encoders have not been \ya{used} for multi-modal recommendation tasks.}}

\pageenlarge{1}
\zxrevise{In this paper, we tackle the above knowledge gap by providing a comprehensive comparison between traditional multi-modal recommendation approaches that do not attempt to align multi-modal item embeddings, and those same approaches when enhanced with \zxrevise{the} LMM encoders across three multi-modal (text+image) recommendation datasets.\footnote{Although our study primarily focuses on visual and textual modalities, the findings may provide insights into the generalisation of such approaches to other modalities, warranting further investigation in future research.} By doing so, we provide both strong conclusions regarding whether \zxrevise{the} LMM encoders should be adopted in state-of-the-art multi-modal recommenders, as well as actionable insights regarding how such encoders should be trained.} 
The primary contributions of this study are three-fold:
$(1)$ \zijunrevise{We systematically \zxrevise{investigate} the integration of two architecturally representative types of LMM encoders, CLIP and VLMo, into five different \ya{recommendation models}. Our \ya{investigation leads to} significant improvements in effectiveness across three distinct recommendation \zxrevise{datasets};}
$(2)$ \ric{We investigate the impact of fine-tuning the CLIP and VLMo} with associated item image and textual descriptions \ric{\zxrevise{from each dataset}, showing that \zxrevise{fine-}tuning leads to increased effectiveness;}
$(3)$ \ric{We compare and contrast} \ya{a} two-step training (i.e., pre-training followed by fine-tuning) with \ya{an} end-to-end training of the encoders.
\zxrevise{Our findings highlight the advantages and implications of using \ya{an} LMM encoder for an improved performance.}

\zxrevise{In summary, we conduct a large-scale empirical investigation addressing 5 dimensions of multi-modal recommendation and their combination, namely recommendation models, multi-modal extractors, training paradigms, datasets and metrics.}
\zxrevise{Our comprehensive evaluation across 480 cases indicates key insights into the effectiveness of the LMM encoders in multi-modal recommendation.}
\zxrevise{Specifically, when integrating pre-trained LMM encoders, we observe significant improvements in 79\% of the 120 tested cases compared to those using modal-specific encoders.}
\zxrevise{Moreover, further significant performance gains are noted when fine-tuning the used three datasets.}
\zxrevise{On the other hand, while a costly end-to-end training results in little performance up-lift for the unified encoder architectures, it significantly benefits the dual-stream LMM encoder. {More generally,} these findings emphasise the importance of establishing \ya{a} deeper modality alignment, facilitated by the LMM encoders, for enhanced representation learning in the multi-modal recommendation task.}

\begin{table}[tb]
\centering
\caption{Comparative analysis of NDCG@20 scores for MMGCL and LATTICE models using different modality inputs across used datasets. V/T are the abbreviations for Visual/Textual, respectively.}
\vspace{-2mm}
\label{tab:analysis}
\begin{adjustbox}{width=\linewidth}
\begin{tabular}{lccccc}
\toprule
\multirow{1}{*}{\textbf{Dataset}} & \multicolumn{1}{c}{Amazon Sports} & \multicolumn{1}{c}{Amazon Clothing} & \multicolumn{1}{c}{Amazon Baby}\\ 
Methods & NDCG@20 & NDCG@20  & NDCG@20\\
\midrule
MMGCL (V\&T) & ${0.0428}$  & ${0.0277}$ & {0.0352} \\
MMGCL (V) & ${0.0420}$  & ${0.0294}$ & {0.0343}  \\
MMGCL (T) & $\textbf{0.0433}$  & \textbf{0.0323} & \textbf{0.0360} \\
\midrule
LATTICE (V\&T) & {0.0424}  & {0.0336} & \textbf{0.0374} \\
LATTICE (V) & {0.0420}  & {0.0343} & {0.0365} \\
LATTICE (T) & \textbf{0.0441}  & \textbf{0.0352} & {0.0372} \\
\bottomrule
\end{tabular}
\end{adjustbox} 
\vspace{-4mm}
\end{table}

\begin{figure*}[tb]
    \begin{subfigure}[t]{0.49\linewidth}
        \includegraphics[trim={0cm 6cm 0cm 6cm},clip,width=1\linewidth]{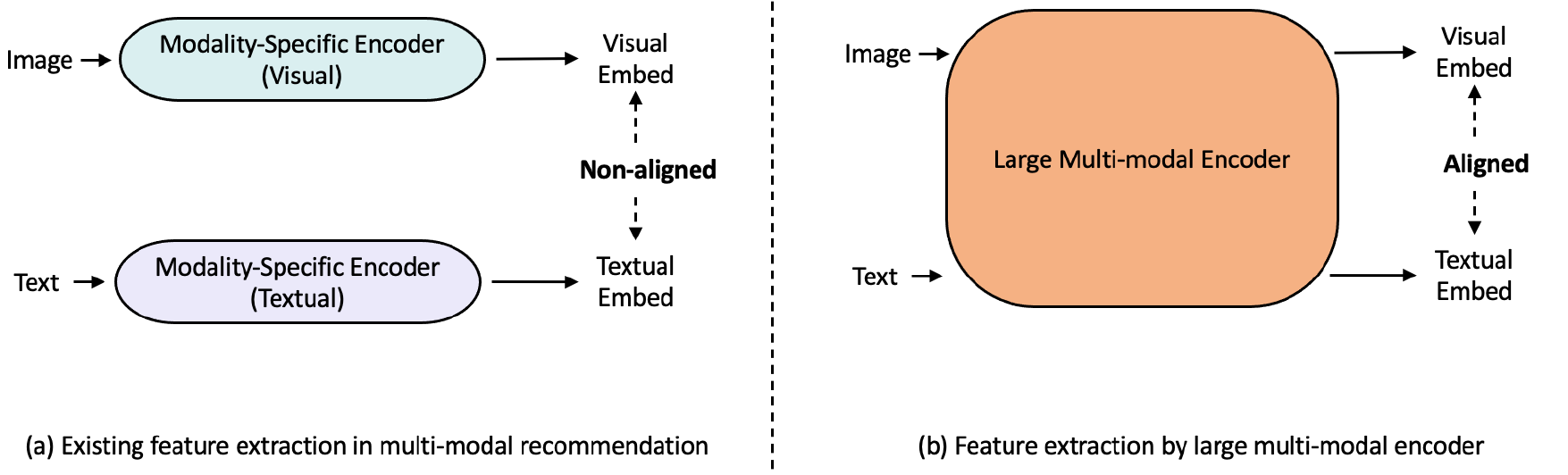}
        \captionsetup{labelformat=empty}
        \caption{Fig.1: An illustration of different feature extraction methods.}
        \label{fig:subfig1}
    \end{subfigure}
    \begin{subfigure}[t]{0.49\linewidth}
        \includegraphics[trim={0cm 11cm 0cm 10cm},clip,width=1\linewidth]{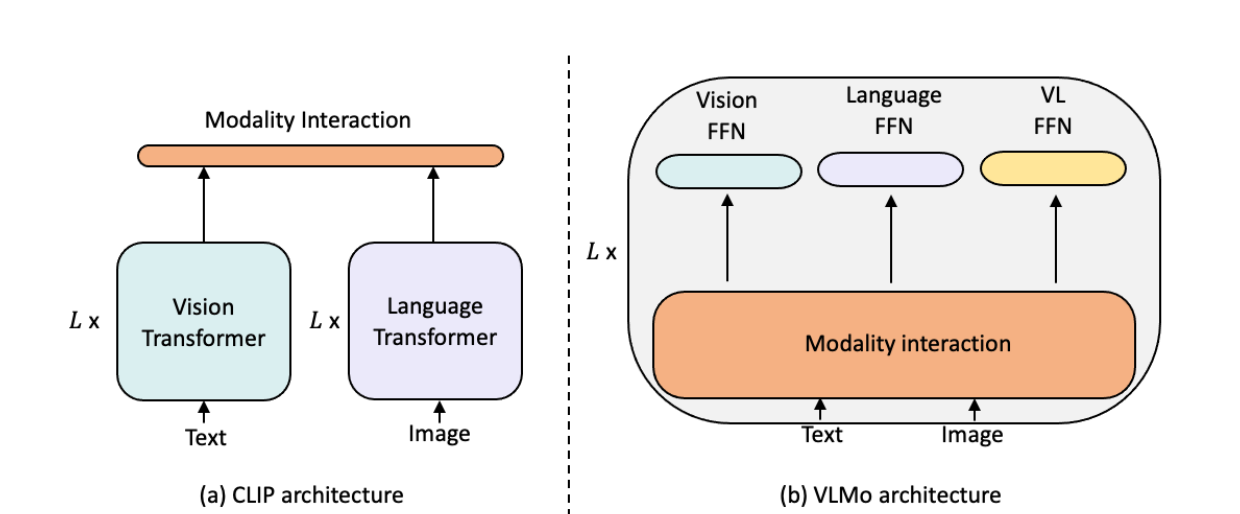}
        \captionsetup{labelformat=empty}
        \caption{Fig.2: The architectures of the large multi-modal encoders.}
        \label{fig:subfig2}
    \end{subfigure}
    \captionsetup{labelformat=simple}
\label{fig}
\vspace{-4mm}
\end{figure*}

\section{Related Work}\label{sec:rwork}
In this section, we discuss related methods and techniques to our conducted study, namely multi-modal recommendation, modality-specific encoders and large multi-modal encoders.

\subsection{Multi-modal Recommendation}\label{sec:mmrec}
\pageenlarge{1}
\looseness -1 Multi-modal recommendation \io{systems aim} to leverage auxiliary multi-modal information, supplementing historical user-item interactions to enhance the recommendation performance~\cite{liu2023multimodal,zhou2023comprehensive,yi2023contrastive}. 
Numerous approaches have been \zixuan{proposed for incorporating multi-modal features into recommendation systems, employing diverse methods to effectively integrate information from different modalities.}
VBPR~\cite{he2016vbpr} is one of the first models to incorporate visual features into recommendation systems by concatenating visual embeddings with ID embeddings \io{in} the item representation. 
MMGCN~\cite{wei2019mmgcn} further advances this approach by injecting high-order semantics into user/item representation learning through several graph convolutional layers. 
This method generates aggregated representations for each modality and combines them using either mean or sum operations, resulting in the final fused representations. 
\zxrevise{Recent advances in multi-modal recommendation have resulted in the emergence of self-supervised learning as a solution, as demonstrated by methods such as MMGCL~\cite{yi2022multi} and SLMRec~\cite{tao2022self}.}
\zxrevise{These models devise augmentations on modality-specific user-item graphs to enhance multi-modal {\em feature alignment}, enabling \ya{to synthesise} information across different modalities for a more coherent representation.}
Another line of \io{approaches} effectively mine item-item structures to enhance item representation learning by capturing the underlying relationships and similarities between items.
For instance, LATTICE~\cite{LATTICE21} constructs item-item graphs for each modality based on the user-item bipartite graphs, performing graph convolutional operations \io{several times} on both item-item graphs and user-item interaction graphs multiple times to obtain more comprehensive and informative user and item representations, which better reflect the complex interactions and dependencies among items and users.
This process contributes to aligning multi-modal features by uncovering latent item-item relationships and associating items with similar modality features.
However, existing methods primarily rely on the concatenation or combination of static and extracted representations from each modality, thereby \zxrevise{only performing a} {\em shallow alignment} for multi-modal fusion, which cannot deeply capture the interrelations among the modalities. 
To the best of our knowledge, there are no existing approaches that perform deep feature alignment for each modality in multi-modal recommendations,
\zixuan{which necessitates comprehensively learning the relationships between modalities to achieve a more effective representation of the multi-modal data.}
In this work, 
we investigate this research direction, with the objective of integrating deep alignment methods like CLIP~\cite{radford2021learning} or VLMo~\cite{bao2022vlmo} as a supplementary component into existing multi-modal recommendation models.

\subsection{Modality-Specific Encoders}\label{sec:enc}
\pageenlarge{1}
\looseness -1 Feature extraction is crucial in multi-modal recommendation because it enables the \craig{identification} of meaningful and discriminative information from various modalities, such as textual, visual, and auditory data~\cite{zhou2023comprehensive}. 
By effectively capturing the intrinsic properties of each modality and their relationships, the recommendation models can better \zixuan{comprehend} and represent the items and users~\cite{tao2022self}.
In the multi-modal recommendation literature, various approaches employ different modality-specific encoders to extract features from raw data. 
VECF~\cite{chen2019personalized} uses the VGG-19 model~\cite{simonyan2014very} for \io{the} pre-segmentation of images, capturing users' attention on different image regions. 
VBPR~\cite{he2016vbpr} extracts visual features from item images using \io{a} pre-trained Deep CNN~\cite{donahue2014decaf}. 
MMGCN~\cite{wei2019mmgcn} employs \zijunrevise{the} ResNet50~\cite{he2016deep} model for visual feature extraction, Sentence2Vector~\cite{le2014distributed} for deriving textual features from micro video descriptions, and VGGish~\cite{gemmeke2017audio} for learning acoustic features. 
LATTICE uses Deep CNN~\cite{donahue2014decaf} and Sentence-Transformer~\cite{reimers2019sentence} for visual and textual feature extraction, respectively.
\zijunrevise{However, employing separate encoders for each modality can result in heterogeneous multi-modal features.
\zxrevise{This means that}
features from different modalities do not inhabit the same semantic space, which can potentially lead to overfitting. This happens as each modality-specific encoder may independently capture noise present in the data, thereby diminishing the model's generalisation capability~\cite{liu2022multi}.}
\zixuan{Moreover, this separation \zxrevise{between uni-modal extractors} can encourage the model to seek shortcuts from uni-modal features on preference scores, rather than effectively leveraging the interdependence between modalities~\cite{LATTICE21}.}
To address the above issues, it is important to ensure consistency among the extracted uni-modal features before inputting them into the recommendation models. Hence, applying a multi-modal extractor that can transform heterogeneous data into a common latent space is a more reasonable approach.
This potentially enhances the performance and generalisation of the recommendation model by effectively capturing the underlying relationships between modalities, allowing for producing more comprehensive item and user representations. 
\zxrevise{In this paper, we leverage the large multi-modal encoders to capture the intrinsic properties of each item's modality and their relationships, thereby resulting in more accurate recommendations.}

\subsection{Large Multi-modal \zixuan{(LMM)} Encoders}\label{sec:lmm}
The remarkable success of transformer-based pre-training in the \io{Natural} Language Processing (NLP) community has led to extensive research on multi-modal pre-training, particularly as various large-scale multi-modal corpora \craig{have emerged}. The self-attention \zijunrevise{mechanism} finds global patterns by examining all word connections, regardless of distance, and effectively captures long-range dependencies without \zijunrevise{using} fixed windows or sequential processing.
This inherent characteristic allows \io{a} transformer to operate in a modality-agnostic manner compatible with various modalities. Recent studies~\cite{rao2022does,wang2023large,yi2023graph} have shown that when pre-trained on large-scale multi-modal corpora, transformer-based models not only significantly outperform their competitors \zixuan{(such as traditional recurrent neural networks and convolutional neural networks)}
across a wide range of multi-modal downstream tasks, \zijunrevise{and} are effective for zero-shot scenarios,
\zixuan{where it is important to be able to generalise to new tasks or domains without any task-specific fine-tuning or additional training.}
In the literature of multi-modal encoders, there are primarily two lines of approaches: \craig{(1)} {\em dual-stream architectures}, such as VSE~\cite{faghri2017vse}, CLIP~\cite{radford2021learning}, ViLBERT~\cite{lu2019vilbert}, which consist of a vision transformer and a language transformer. The vision transformer processes images, while the language transformer handles textual data. Both encoders generate embeddings for their respective inputs, which are then aligned using several fusion layers;
(2) {\em unified architectures}, such as VLMo~\cite{bao2022vlmo}, which jointly processes multi-modal data into a Mixture-of-Modality-Experts (MoME) Transformer to obtain contextualised representations and align the visual and language feature vectors.
The MoME Transformer is used to encode different modalities, with a mixture of modality experts replacing the feed-forward network of a standard Transformer. Each MoME Transformer block captures modality-specific information by switching to a different modality expert and employs multi-head self-attention (MSA) shared across modalities to align visual and text content.
As discussed in Section~\ref{sec:mmrec} and Section~\ref{sec:enc},
feature alignment and feature extraction are important research points in multi-modal recommendation systems. 
\zxrevise{These extraction and alignment processes
can capture and exploit the relationships between different modalities and generate more effective representations for downstream tasks.}
\zxrevise{Despite the significant benefits offered by \zxrevise{the} LMM encoders, their integration into recommendation systems has not been investigated extensively to-date. Hence,}
in this work, we choose two representative LMM encoders, CLIP and VLMo, each representing distinct architectural approaches to multi-modal encoders, for the purpose of extracting and aligning multiple modalities in the context of the recommendation task under various training paradigm settings.

\section{Probing Large Multi-modal Encoders in Recommendation}
\looseness -1 In light of the advantages of \zxrevise{CLIP and VLMo} for multi-modal representation learning, as discussed in Section~\ref{sec:rwork}, we
detail the settings in which we extend the use of \zxrevise{CLIP and VLMo} for the recommendation task.
First, we describe the process of using multi-modal embeddings, \zxrevise{obtained from \zxrevise{CLIP and VLMo}}, to initialise the user/item embeddings in the existing recommendation models.
\zxrevise{Then, we \ya{describe} the method for fine-tuning \zxrevise{CLIP and VLMo} using \ya{the} recommendation datasets and \zxrevise{illustrate} the integration of \zxrevise{CLIP and VLMo} with \ya{the} existing recommendation models in an end-to-end approach.}

\subsection{Multi-modal Encoding through LMMs}
\looseness -1 \zxrevise{In this section, we introduce how \zxrevise{CLIP and VLMo} encode multi-modal item representations from the raw data:
\noindent (1) \textbf{CLIP:} For CLIP, raw images and texts are encoded into image and text vector representations. CLIP leverages the Vision Transformer (ViT) architecture~\cite{dosovitskiy2020image} to process image representations by dividing the input image into non-overlapping patches, flattening them into vectors, and linearly projecting them to create patch embeddings. 
Text representations are generated using the GPT-2~\cite{radford2019language} model, after tokenising the raw text input using byte pair encoding (BPE) and adding positional embeddings.
\noindent (2) \textbf{VLMo:} } \zijunrevise{Unlike CLIP, VLMo operates as a unified multi-modal model. It processes the concatenation of image and text inputs as a single unit and produces \ya{the} corresponding vector embeddings. Image inputs are created by splitting images into patches, flattening these patches, and then linearly projecting them to form patch embeddings. A \zxrevise{learned} special token [I\_CLS] is added to the sequence, alongside \ya{the} position and type embeddings. \ya{The text} input comprises tokens generated from raw text using a BERT \cmw{tokeniser}, with the addition of a start-of-sequence token ([T\_CLS]) and a boundary token ([T\_SEP]). Just like the image input, the final text input is a composite of word, position, and type embeddings. Following this, the MoME Transformer is deployed to encode different modalities, with \zxrevise{the} language and visual \zxrevise{expert} \zxrevise{components}~\cite{bao2022vlmo} respectively extracting modality-specific information to produce \ya{the} textual and visual embeddings.}

To illustrate how the multi-modal embeddings of CLIP and VLMo can be used as initial item embeddings in multi-modal recommendation models, we use the VLMo-Base Plus \zijunrevise{model} variant as an example.
The resulting text embeddings for all items from the VLMo-Base Plus encoder have a shape of [item number, text\textunderscore token\textunderscore length+2, 544], where each item comprises the number of raw text tokens along with [CLS] and [SEP] tokens. As for image embeddings, each item has an embedding shape of [197, 544].
We obtain the item visual and text embeddings by selecting the 544-dimensional vector corresponding to the [CLS] token for each item embedding which should encapsulate rich, high-level information in each modality.

\subsection{Training Strategies of the LMM Encoders}
\pageenlarge{1}
\looseness -1 In this section, we present various training paradigms for incorporating \zxrevise{the} \zixuan{LMM} encoders into existing multi-modal recommendation models. We discuss the optimisation objectives used to tune both the \zixuan{LMM} encoders and the recommendation models. We aim to identify \zixuan{the best} 
training paradigm that facilitates the effective integration of \zxrevise{the LMM} encoders, leveraging their strengths to enhance the performance of existing recommendation models:
$(1)$ \textbf{Two-stage training}
\looseness -1 involves first fine-tuning the \zxrevise{LMM} encoders on item images and texts, and subsequently using the derived item embeddings as initial embeddings for each modality in \ya{the} existing recommendation models. \zijunrevise{This \zxrevise{fine-tuning} process
is designed to enhance the adaptability and performance of \zxrevise{the LMM} encoders in the context of recommendation} \io{scenarios}.
Specifically, we tune both CLIP and VLMo using \io{their} associated loss functions, which include ITC loss, MLM loss, and ITM loss~\cite{radford2021learning,bao2022vlmo};
$(2)$ \textbf{End-to-end training}
typically needs the seamless integration of \ya{the} \zixuan{LMM} encoders so as to jointly optimise both \io{the} \zixuan{LMM} encoders and \io{the} existing recommendation models with the recommendation loss.
\zixuan{This integration must account for the different types of losses used by each model, such as the BPR loss~\cite{rendle2009bpr} and \ya{the} contrastive loss~\cite{wu2021self}.}
\zixuan{Although we acknowledge that some recommendation models (e.g. LATTICE) facilitate feature alignment and may have overlapping concepts with CLIP and VLMo, our investigation focuses on the impact of integrating these \zixuan{LMM} encoders into various recommendation architectures.}
Specifically, we \zixuan{leverage the original implementations provided by the authors for
the \craig{CLIP and VLMo} architectures}
to process the item images and their corresponding text to extract \ya{the} visual and \zijunrevise{textual} embeddings.
\zixuan{These embeddings are then used to initialise item representations in the downstream recommendation models.}
Algorithm~\ref{fig:pseudo} (above) \zixuan{presents the pseudo-code for end-to-end training.}


\begin{algorithm}[t]
\caption{End-to-end training}
\label{fig:pseudo}
  \scriptsize
\begin{algorithmic}[1]
    \State \textbf{Step 1}: Load data
    \State \hspace{0.5cm} \textit{data\_loader} $\leftarrow$ DataLoader(\textit{dataset\_path}, \textit{raw\_images}, \textit{concatenated\_texts(title, description, brand, categorical\_info)})
    \State \textbf{Step 2}: Initialise and load pre-trained weights for CLIP/VLMo encoder
    \State \hspace{0.5cm} \textit{clip/vlmo} $\leftarrow$ CLIP()/VLMo()
    \State \hspace{0.5cm} \textit{clip/vlmo.load\_pretrained\_weights()}
    \State \textbf{Step 3}: Generate embeddings with CLIP/VLMo encoder
    \State \hspace{0.5cm} \textit{image\_embeddings, text\_embeddings, image\_text\_embeddings} $\leftarrow$ \textit{clip/vlmo.generate\_embeddings(data\_loader)}
    \State \textbf{Step 4}: Integrate embeddings into the recommendation model
    \State \hspace{0.5cm} \textit{rec\_model} $\leftarrow$ REC(\textit{image\_embeddings, text\_embeddings, image\_text\_embeddings})
    \For{epoch in range(num\_epochs)}
        \State \hspace{0.5cm} \textbf{Step 5}: End-to-end training
        \State \hspace{1cm} \# Forward pass
        \State \hspace{1cm} \textit{user\_item\_scores} $\leftarrow$ \textit{rec\_model.forward()}
        \State \hspace{1cm} \# Compute loss
        \State \hspace{1cm} \textit{loss} $\leftarrow$ compute\_loss(\textit{user\_item\_\_scores}, \textit{ground\_truth})
        \State \hspace{1cm} \# Backward pass and optimisation
        \State \hspace{1cm} \textit{optimiser.zero\_grad()}
        \State \hspace{1cm} \textit{loss.backward()}
        \State \hspace{1cm} \textit{optimiser.step()}
        \State \hspace{1cm} \# Update model with new embeddings
        \State \hspace{1cm} \textit{rec\_model.update(image\_embeddings, text\_embeddings, image\_text\_embeddings)}
        \State \hspace{1cm} \# Evaluate and print performance metrics
        \State \hspace{1cm} \textit{evaluate\_and\_print\_metrics(epoch, rec\_model)}
    \EndFor
\end{algorithmic}
\end{algorithm}\vspace{-1.5mm}

\section{Experiments}
In this section, we conduct experiments to \zxrevise{examine} the effectiveness of \zxrevise{CLIP and VLMo} on three public datasets, in comparison to five existing state-of-the-art multi-modal recommendation models.
To \zxrevise{evaluate} the effectiveness of \zxrevise{CLIP and VLMo}, we conduct experiments to answer the following three research questions:

\noindent \textbf{RQ1}: 
Do \zxrevise{the} pre-trained \zxrevise{CLIP and VLMo} outperform existing extractors in multi-modal recommendation? \\
\noindent \textbf{RQ2}:
\zxrevise{Does the strategy of fine-tuning CLIP and VLMo on \ya{the} raw images and \zxrevise{the} text of recommendation items, or using an end-to-end training paradigm, produce additional performance benefits for \ya{the} multi-modal recommendation models?}
\\
\noindent \textbf{RQ3}: 
Do \zxrevise{CLIP and VLMo} effectively align different modalities in multi-modal recommendation systems?


\subsection{Experimental Settings}
\pageenlarge{1}
\subsubsection{Datasets}
\zijunrevise{To assess the performance of CLIP and VLMo in the recommendation task, we \ya{carry out} experiments on three \ya{widely used} Amazon Review datasets: Sports and Outdoors (abbreviated as Sports), Clothing, Shoes and Jewelry (referred to as Clothing), and Baby. Sourced from the Amazon Review datasets repository, these datasets are not only frequently used benchmarks in \ya{the} recommendation systems \ya{literature} but \ya{they} also offer rich multi-modal data~\cite{liu2023multimodal,zhou2023comprehensive}. These datasets provide features such as item image URLs and \ya{their} descriptions, enabling \ya{a} robust and representative evaluation of multi-modal recommendation performance~\cite{LATTICE21}.}
Processing the datasets is a crucial step in multi-modal recommendation, \ya{since} we need to handle unprocessed datasets containing extensive information from different modalities. 
Following the dataset processing \io{protocol}, widely used in previous works~\cite{LATTICE21, zhou2023comprehensive}, we transform the ratings into binary values of 0 or 1, indicating whether the user has rated the item. 
In line with previous works~\cite{LATTICE21}, we filter out users and items with more than 5 interactions in a given dataset. 
Unlike existing approaches~\cite{he2016vbpr, wei2019mmgcn, yi2022multi, LATTICE21} that \io{use} pre-extracted features within the datasets, we download \io{the} raw images from \io{the} item URLs and encode them with \zxrevise{the} \zixuan{LMM} encoders, \io{instead of} using pre-extracted 4096-dimensional visual features of items~\cite{ni2019justifying}. 
For \io{the} textual features, we employ the title, description, brand, and categorical information of items and \io{also} encode them with \zxrevise{the} \zixuan{LMM} encoders. 
Contrasting with existing approaches that use \zixuan{Sentence-Transformer} to extract 384-dimensional textual embeddings~\cite{LATTICE21}, we use CLIP and VLMo to \zijun{encode} the raw text of items into 768 and 544  dimensions, respectively. 
The exact statistics of the used datasets are presented in Table~\ref{tab:stat_all}. 

\begin{table}[tb]
\centering
\caption{Statistics of the used Amazon datasets.}
\vspace{-2mm}
\resizebox{0.48\textwidth}{!}{
\begin{tabular}{l|ccc}
     \hline
     & Sports & Clothing & Baby\\
     \hhline{====}
     Users &  35,598 &  39,387 &  19,445 \\
     Items &  18,287 & 22,499 & 7,037\\
     Interactions &295,366 & 271,001 &  160,522\\
     \hline
     \zxrevise{Interaction} Density &0.00045 &0.00030 & 0.00012\\    
     \hline
    CNN/CLIP/VLMo Visual Dimension & 4096/768/544 & 4096/768/544  & 4096/768/544\\
    Sentence-Transformer/CLIP/VLMo Textual Dimension & 384/768/544 & 384/768/544 & 384/768/544\\
    CNN+Sentence-Transformer/CLIP/VLMo Parameters (million) & 170/151/167 & 170/151/167 & 170/151/167\\
     \hline
\end{tabular}}
\label{tab:stat_all}
\vspace{-6mm}
\end{table}

\subsubsection{Evaluation Protocols}
\looseness -1 \io{Similar to} the evaluation setting in ~\cite{LATTICE21, zhou2023comprehensive}, we randomly split the datasets into training, validation, and testing sets with an 8:1:1 ratio. 
To perform negative sampling for each user, we \zxrevise{sample} items that have no prior interactions with the user from the history of observed user-item interactions.
We use two commonly used evaluation metrics, \io{namely} Recall@K and NDCG@K, to evaluate the performance of top-$K$ {recommendation}. 
We follow~\cite{LATTICE21} in setting K = 10, 20 and report the average performance achieved for all users in the testing set. 
Following the settings in ~\cite{wei2019mmgcn, zhou2023comprehensive}, we \zixuan{use an} all-rank
item evaluation strategy is used to measure the used metrics.
We use the Adam~\cite{kingma2014adam} optimiser in both the \zixuan{LMM} enhanced models and the {five} baseline models.
We apply \io{an} early-stopping strategy during training, terminating the training when the validation loss does not decrease for 50 epochs.

\subsubsection{Baselines}\label{sec:baseline}
\pageenlarge{1}

To examine the effectiveness of \zxrevise{the} \zixuan{LMM} encoders, we compare 
the performance of recommendation models using pre-trained modality-specific encoders—employing CNN and Sentence-Transformer independently for each modality—with that of \zxrevise{the} pre-trained \zixuan{LMM} encoders like \craig{CLIP and VLMo},
which jointly extract information from both modalities. 
\zixuan{In this paper, we choose five state-of-the-art multi-modal recommendation models, as follows:}

\begin{itemize}
    \item \textbf{VBPR~\cite{he2016vbpr}:} This model integrates multimedia features into the matrix factorisation for recommendation. Specifically, it incorporates visual features into the matrix decomposition. 
    
    \item \textbf{MMGCN~\cite{wei2019mmgcn}:} This model employs graph convolutional networks (GCN) to propagate modality-specific embeddings and capture modality-related user preferences for multi-modal recommendation. The final user and item representations are generated by combining the learned representations from each modality, resulting in improved multi-modal recommendations. 
\zixuan{In our experiments, we differentiate MMGCN from VBPR by not only incorporating visual features but also integrating textual features derived from CLIP and VLMo.}
    
    \item \textbf{MMGCL~\cite{yi2022multi}:} This is a self-supervised graph learning model \io{that} leverages modality edge dropout and modality masking to learn complex user preferences. Furthermore, it introduces a novel negative sampling technique to learn the correlation between multiple modalities and performs multi-task learning by combining both Bayesian personalised ranking (BPR) and self-supervised loss. 
    
    \item \textbf{SLMRec~\cite{tao2022self}:} This is also a self-supervised graph learning model. It emphasises the importance of individual modalities by creating fine and coarse spaces to align features across modalities, thereby enhancing consistency for improved fusion. 
By treating each modality as a distinct feature, this model leverages self-supervised learning to generate supervised signals by contrasting different item embedding via augmentation. 
Different from MMGCL, which uses multi-task loss, SLMRec only leverages \io{a} self-supervised learning loss as the main loss.

    \item \textbf{LATTICE~\cite{LATTICE21}:} This model constructs item-item graphs for each modality based on the user-item bipartite graph and subsequently aggregates them to generate latent item graphs. It focuses on mining latent semantic structures between items by learning item-item graphs derived from their multi-modal features. The model then performs graph convolutional operations on both \io{the} item-item graphs and user-item interaction graphs to obtain user and item representations, ultimately identifying latent item-item relations and connecting items with similar modality features. 
\end{itemize}

\subsubsection{\craig{Model Checkpoints and} Hyperparameter Settings}
\pageenlarge{1}
\looseness -1 \zijun{All used baselines (VBPR\footnote{\url{https://github.com/DevilEEE/VBPR}}, MMGCN\footnote{\url{https://github.com/weiyinwei/MMGCN}}, MMGCL\footnote{\url{https://github.com/zxy-ml84/MMGCL}}, SLMRec\footnote{\url{https://github.com/zltao/SLMRec}}, LATTICE\footnote{\url{https://github.com/CRIPAC-DIG/LATTICE}}) and \io{the} \zixuan{LMM} encoders (\zixuan{CLIP ViT-B/16}\footnote{\url{https://github.com/openai/CLIP}}, \zixuan{VLMo-Base Plus}\footnote{\url{https://github.com/microsoft/unilm/tree/master/vlmo}}) are implemented with PyTorch and trained on a GPU A6000 with 48GB of memory.}
\zijunrevise{To facilitate a comparison of the impact of \ya{the} LMM encoders on \ya{the} recommendation effectiveness, we make deliberate choices for \ya{the} CLIP and VLMo variants based on their reported performances in the \ya{literature}. Specifically, for CLIP, we opt for the ViT-B/16 variant as the image encoder, motivated by its superior performance in image tasks when compared to other CNN image encoders within the CLIP model framework~\cite{radford2021learning}. For VLMo, our choice is the VLMo-Base Plus model, similarly driven by its demonstrated effectiveness~\cite{bao2022vlmo}. Both models have comparable numbers of parameters (151 vs. 167 million), ensuring a fair comparison. The architectural differences between CLIP's ViT-B/16 and VLMo-Base are depicted in Fig.~2, \ya{which offers} a visual juxtaposition of their structures.
}
\craig{We use the authors' original code for VLMo and CLIP in our experiments, but note that their code for} VLMo is implemented with the PyTorch Lightning framework. We converted it into pure PyTorch code without using the PyTorch Lightning library. Our motivations for converting the original code from PyTorch Lightning to pure PyTorch include greater customisation, compatibility, and performance considerations. By using pure PyTorch, we gain more control and flexibility, allowing us to tailor the code to specific requirements, which is beneficial for future research needs. 
\zijunrevise{While optimising the LMM encoders, we observe a marked acceleration in training speed upon transitioning from PyTorch Lightning to pure PyTorch. In our quest to pinpoint the most effective hyperparameters, we undertake extensive parameter searches for each recommendation dataset, using metrics from the validation set. These evaluations are performed during both the fine-tuning phase and the end-to-end training. Specifically, we experimented with learning rates spanning $\left \{ 1\mathrm{e}{-5}, 3\mathrm{e}{-5}, 5\mathrm{e}{-5}, 1\mathrm{e}{-4}, 1\mathrm{e}{-3} \right \}$ and varied batch sizes, including  $\left \{32, 64, 128, 256, 1024 \right \}$.}
\zijunrevise{The grid search procedure is conducted in accordance with the available code of MMRec\footnote{\url{https://github.com/enoche/MMRec}} and is applied to all model variants we \ya{evaluated} in the experiments.}


\subsection{Pre-trained Modality-specific Encoder vs. Pre-trained LMM Encoders (RQ1)}\label{sec:rq1}

\begin{table*}[tb]
\centering
\vspace{-6mm}
\caption{Experimental results between using modality-specific encoders and using \zxrevise{the} \zixuan{LMM} encoders in multi-modal recommenders. The best performance of each model is highlighted in bold. $^{*}$ denotes a significant difference compared to the result of baselines using the paired t-test with $p<0.05$. }
\vspace{-2mm}
\label{tab:comp_base}
\begin{adjustbox}{width=\linewidth}
\begin{tabular}{lcccccccccccccc}
\toprule
\multirow{1}{*}{\textbf{Dataset}} & \multicolumn{4}{c}{Amazon Sports} & \multicolumn{4}{c}{Amazon Clothing} & \multicolumn{4}{c}{Amazon Baby}\\ 
\cmidrule(lr){1-1} \cmidrule(lr){2-5} \cmidrule(lr){6-9} \cmidrule(lr){10-13}
Methods & Recall@10 & Recall@20 & NDCG@10 & NDCG@20 & Recall@10 & Recall@20 & NDCG@10 & NDCG@20  & Recall@10 & Recall@20 & NDCG@10 & NDCG@20\\
\midrule
VBPR & ${0.0509}$ & ${0.0771}$ &  ${0.0280}$ & ${0.0349}$  &  ${0.0409}$ & ${0.0611}$ & ${0.0226}$ & ${0.0277}$ & ${0.0479}$ & ${0.0740}$ & ${0.0262}$ & ${0.0329}$\\
VBPR$_{VLMo}$ & \textbf{{0.0562}$^{*}$} & $\textbf{0.0848}^{*}$ &  $\textbf{0.0299}^{*}$ & $\textbf{0.0373}^{*}$  &  $0.0399^{*}$ &  $0.0577^{*}$ &  $0.0221$ &  $0.0266^{*}$ &  $\textbf{0.0497}^{*}$ &  $\textbf{0.0774}^{*}$ &  $\textbf{0.0274}^{*}$ &  $\textbf{0.0346}^{*}$ \\
VBPR$_{CLIP}$ & ${0.0536}^{*}$ & ${0.0802}^{*}$ &  ${0.0288}^{*}$ & ${0.0357}^{*}$  & $\textbf{0.0435}^{*}$ & $\textbf{0.0675}^{*}$ & $\textbf{0.0235}^{*}$ & $\textbf{0.0296}^{*}$ & ${0.0487}^{*}$ & ${0.0762}^{*}$ & ${0.0265}^{*}$ & ${0.0336}^{*}$\\
\midrule
MMGCN & ${0.0290}$  & ${0.0475}$  &  ${0.0154}$ & ${0.0201}$ & ${0.0151}$ &  ${0.0246}$ & {0.0077} & {0.0100} & \textbf{0.0391} & \textbf{0.0642} & \textbf{0.0201} & \textbf{0.0266} \\
MMGCN$_{VLMo}$ & ${0.0295}^{*}$  & ${0.0484}^{*}$  &  ${0.0161}^{*}$ & ${0.0208}^{*}$ & ${0.0156}^{*}$ & ${0.0252}^{*}$ & ${0.0086}^{*}$ & $\textbf{0.0104}^{*}$ & ${0.0336}$$^{*}$ & ${0.0538}$$^{*}$ & ${0.0172}$$^{*}$ & ${0.0224}$$^{*}$ \\
MMGCN$_{CLIP}$ & $\textbf{0.0312}^{*}$  & $\textbf{0.0494}^{*}$  &  $\textbf{0.0165}^{*}$ & $\textbf{0.0216}^{*}$ & $\textbf{0.0163}^{*}$ & $\textbf{0.0253}^{*}$ & $\textbf{0.0090}^{*}$ & ${0.0101}$ & ${0.0376}$$^{*}$ & ${0.0606}$$^{*}$ & ${0.0198}$ & ${0.0257}$ \\
\midrule
MMGCL & ${0.0617}$  & ${0.0913}$ & {0.0351} & {0.0428} & {0.0410}& {0.0607}& {0.0227}& {0.0277}& {0.0521}& {0.0790}& {0.0283}& {0.0352} \\
MMGCL$_{VLMo}$ & $\textbf{0.0649}^{*}$  & ${0.0941}^{*}$ & \textbf{0.0371}$^{*}$ & {0.0446}$^{*}$ & ${0.0446}^{*}$ & ${0.0648}^{*}$ & ${0.0247}^{*}$ & ${0.0299}^{*}$ & $\textbf{0.0540}^{*}$ & $\textbf{0.0822}^{*}$ & $\textbf{0.0298}^{*}$ & $\textbf{0.0370}^{*}$ \\
MMGCL$_{CLIP}$ & ${0.0647}^{*}$  & $\textbf{0.0959}^{*}$ & {0.0367}$^{*}$ & \textbf{0.0447}$^{*}$ & $\textbf{0.0478}^{*}$ & $\textbf{0.0703}^{*}$ & $\textbf{0.0264}^{*}$ & $\textbf{0.0321}^{*}$ & ${0.0538}^{*}$ & ${0.0804}^{*}$ & ${0.0293}^{*}$ & ${0.0362}^{*}$ \\
\midrule
SLMRec & {0.0605}  & {0.0901} & ${0.0341}$ & {0.0417} & {0.0430} & {0.0623} & {0.0238} & {0.0287} & {0.0527} & {0.0810} & {0.0288} & {0.0361} \\
SLMRec$_{VLMo}$ & {0.0615}$^{*}$  & \textbf{0.0919}$^{*}$ & ${0.0348}^{*}$ & ${0.0426}^{*}$ & {0.0450}$^{*}$ & {0.0653}$^{*}$ & \textbf{0.0252}$^{*}$ & \textbf{0.0304}$^{*}$ & {0.0561}$^{*}$ & \textbf{0.0823}$^{*}$ & {0.0302}$^{*}$ & {0.0374}$^{*}$ \\
SLMRec$_{CLIP}$ & \textbf{0.0630}$^{*}$  & {0.0909}$^{*}$ & $\textbf{0.0354}^{*}$ & $\textbf{0.0427}^{*}$ & \textbf{0.0451}$^{*}$ & \textbf{0.0671}$^{*}$ & {0.0249}$^{*}$ & \textbf{0.0304}$^{*}$ & \textbf{0.0568}$^{*}$ & {0.0821}$^{*}$ & \textbf{0.0310}$^{*}$ & \textbf{0.0381}$^{*}$ \\
\midrule
LATTICE & \textbf{0.0633}  & \textbf{0.0944} & \textbf{0.0334} & \textbf{0.0424} & \textbf{0.0484} & \textbf{0.0704} & \textbf{0.0280} & \textbf{0.0336} & \textbf{0.0539} & \textbf{0.0860} & \textbf{0.0291} & \textbf{0.0374} \\
LATTICE$_{VLMo}$ & {0.0472}$^{*}$  & {0.0758}$^{*}$ & {0.0242}$^{*}$ & {0.0312}$^{*}$ & {0.0344}$^{*}$ & {0.0512}$^{*}$ & {0.0177}$^{*}$ & {0.0209}$^{*}$ & {0.0502}$^{*}$ & {0.0817}$^{*}$ & {0.0264}$^{*}$ & {0.0348}$^{*}$ \\
LATTICE$_{CLIP}$ & {0.0500}$^{*}$  & {0.0785}$^{*}$ & {0.0259$^{*}$} & {0.0332}$^{*}$ & {0.0396}$^{*}$ & {0.0593}$^{*}$ & {0.0211}$^{*}$ & {0.0263}$^{*}$ & {0.0524}$^{*}$ & {0.0830}$^{*}$ & {0.0275}$^{*}$ & {0.0354}$^{*}$ \\
\bottomrule
\end{tabular}
\end{adjustbox} 
\end{table*}

\zixuan{As discussed in Section~\ref{sec:baseline}, to ensure a fair comparison, we primarily focus on the results obtained from recommendation models that use pre-trained modality-specific encoders. These models employ CNN and Sentence-Transformer independently for each modality, providing a consistent baseline for evaluating the impact of incorporating CLIP and VLMo.}
These results are compared with those using CLIP ViT-B/16 and VLMo-Base Plus, which jointly extract information from both modalities. 
To evaluate the statistical significance of 
\zixuan{performance differences between the five selected multi-modal recommendation models with and without the integration of CLIP/VLMo,}
we use the paired t-test ($p<0.05$).
Table~\ref{tab:comp_base} presents the results of our conducted experiments \zxrevise{across 120 cases}, comparing the performance of recommendation models using pre-trained modality-specific encoders (VBPR, MMGCN, MMGCL, SLMRec, and LATTICE) with those employing \zxrevise{the} pre-trained \zixuan{LMM} encoders (VBPR$_{CLIP/VLMo}$, MMGCN$_{CLIP/VLMo}$, MMGCL$_{CLIP/VLMo}$, 

\noindent SLMRec$_{CLIP/VLMo}$, LATTICE$_{CLIP/VLMo}$) in the context of \io{a} multi-modal recommendation task.
From the table, we observe that for all three used datasets, \zxrevise{79\% of the cases tested with} \ya{the} recommendation models (VBPR, MMGCN, MMGCL, SLMRec), using CLIP/VLMo as encoders, significantly outperform \zxrevise{the models using} the original modality-specific encoders. 
This \io{observation} demonstrates the effectiveness of using \zxrevise{the} \zixuan{LMM} encoders as feature extractors, which enables collaborative multi-modal feature generation and mitigates the issue of heterogeneity between visual and textual modalities.
\zixuan{We now focus on comparing the LATTICE variants, a model that exhibits distinct trends among the five baselines. Indeed, we observe that LATTICE performs generally better than LATTICE$_{CLIP}$ and LATTICE$_{VLMo}$ on all three used datasets. This contrasts with the other baselines models, which are generally improved by CLIP and VLMO,}
and \io{suggests} a \zixuan{possible} discrepancy between CLIP/VLMo and LATTICE in terms of feature alignment. \craig{Recall from Section~\ref{sec:mmrec} that LATTICE constructs} item-item graphs based on \craig{the} semantic similarities across different modalities. This objective is conceptually in conflict with that of CLIP/VLMo, which generates deeply aligned features that could lead to a denser item-item graph in LATTICE. Consequently, the features extracted by CLIP/VLMo may result in inadequate item-item graphs, leading to a decline in performance.
On the other hand, as observed in Table~\ref{tab:comp_base}, the recommendation models employing CLIP as an encoder and those using VLMo as an encoder exhibit similar performances 
for all three datasets, with the exception of CLIP outperforming VLMo on \io{the} Amazon Clothing dataset. 
\zixuan{To further determine the optimal multi-modal encoder with different settings, we continue our investigation in the following experiments involving fine-tuning and end-to-end training paradigms.}
\zijunrevise{Overall, \ya{to answer} RQ1, we \ya{conducted} a large-scale empirical investigation \zxrevise{of the} \zxrevise{pre-training} setup. \ya{Our} exploration \ya{addressed} five dimensions of multi-modal recommendation and their combinations: recommendation models, multi-modal extractors, training paradigms, datasets, and metrics. In total, 120 cases \ya{were} examined. \ya{From the experiments, we} conclude that \ya{the} pre-trained LMM encoders are more effective at extracting and aligning visual and textual features from raw images and text, especially when compared to methods using CNN and Sentence-Transformer. \ya{Indeed, we observe a} significant improvement in performance \ya{in} 79\% of the tested cases.} 

\subsection{Fine-tuning \& End-to-end (RQ2)}\label{sec:rq2}
\begin{table*}[tb]
\centering
\caption{Experimental results \yo{comparing} using the pre-trained \& fine-tuned LMM encoders and using the end-to-end LMM encoders in multi-modal recommenders. The best performance of each model is highlighted in bold. $^{*}$ denotes a significant difference compared \io{to the} best result using the paired t-test with the Holm-Bonferroni correction for $p<0.05$. PT/FT/ETE are the abbreviations for Pre-Training, Fine-Tuning and End-To-End, respectively.}
\vspace{-2mm}
\label{tab:ft}
\begin{adjustbox}{width=\linewidth}
\begin{tabular}{lcccccccccccccc}
\toprule
\multirow{1}{*}{\textbf{Dataset}} & \multicolumn{4}{c}{Amazon Sports} & \multicolumn{4}{c}{Amazon Clothing} & \multicolumn{4}{c}{Amazon Baby}\\ 
\cmidrule(lr){1-1} \cmidrule(lr){2-5} \cmidrule(lr){6-9} \cmidrule(lr){10-13}
Methods & Recall@10 & Recall@20 & NDCG@10 & NDCG@20 & Recall@10 & Recall@20 & NDCG@10 & NDCG@20  & Recall@10 & Recall@20 & NDCG@10 & NDCG@20\\
\midrule
VBPR$_{VLMo-PT}$ & {{0.0562}$^{*}$} & ${0.0848}^{*}$ &  ${0.0299}^{*}$ & ${0.0373}^{*}$  &  $0.0399^{*}$ &  $0.0577^{*}$ &  ${0.0221}^{*}$ &  $0.0266^{*}$ &  ${0.0497}^{*}$ &  ${0.0774}^{*}$ &  ${0.0274}^{*}$ &  ${0.0346}^{*}$ \\
VBPR$_{VLMo-FT}$ & \textbf{{0.0593}} & $\textbf{0.0877}$ &  $\textbf{0.0321}$ & $\textbf{0.0395}$  &  $\textbf{0.0418}$ &  $\textbf{0.0614}$ &  $\textbf{0.0230}$ &  $\textbf{0.0280}$ &  $\textbf{0.0523}$ &  $\textbf{0.0804}$ &  $\textbf{0.0283}$ &  $\textbf{0.0356}$ \\
VBPR$_{VLMo-ETE}$ & {{0.0485}$^{*}$} & ${0.0706}^{*}$ &  ${0.0271}^{*}$ & ${0.0329}^{*}$ 
&  ${0.0286}^{*}$ &  ${0.0445}^{*}$ &  ${0.0159}^{*}$ &  ${0.0199}^{*}$ 
&  ${0.0518}$ &  ${0.0799}$ &  ${0.0273}^{*}$ &  ${0.0346}$ \\
\midrule
VBPR$_{CLIP-PT}$ & ${0.0536}^{*}$ & ${0.0802}^{*}$ &  ${0.0288}^{*}$ & ${0.0357}^{*}$ & ${0.0435}^{*}$ & ${0.0675}^{*}$ & ${0.0235}^{*}$ & ${0.0296}^{*}$ & ${0.0487}^{*}$ & ${0.0762}^{*}$ & ${0.0265}$ & ${0.0336}^{*}$\\
VBPR$_{CLIP-FT}$ & ${0.0546}^{*}$ & ${0.0818}^{*}$ &  ${0.0294}^{*}$ & ${0.0364}^{*}$ & $\textbf{0.0468}$ & $\textbf{0.0715}$ & $\textbf{0.0250}$ & $\textbf{0.0313}$ & $\textbf{0.0506}$ & $\textbf{0.0790}$ & $\textbf{0.0271}$ & $\textbf{0.0344}$\\
VBPR$_{CLIP-ETE}$ & $\textbf{0.0594}$ & $\textbf{0.0892}$ &  $\textbf{0.0322}$ & $\textbf{0.0395}$ 
& ${0.0433}^{*}$ & ${0.0674}^{*}$ & ${0.0234}^{*}$ & ${0.0295}^{*}$ 
& ${0.0490}^{*}$ & ${0.0766}$ & ${0.0269}$ & ${0.0340}$\\
\midrule
MMGCN$_{VLMo-PT}$ & ${0.0295}^{*}$  & ${0.0484}^{*}$  &  ${0.0161}^{*}$ & ${0.0208}^{*}$ & ${0.0156}^{*}$ & ${0.0252}^{*}$ & ${0.0086}^{*}$ & ${0.0104}^{*}$ & ${0.0336}^{*}$ & ${0.0538}^{*}$ & ${0.0172}^{*}$ & ${0.0224}^{*}$ \\
MMGCN$_{VLMo-FT}$ & ${0.0315}$  & ${0.0511}^{*}$  &  ${0.0168}$ & ${0.0218}$ & ${0.0165}^{*}$ & ${0.0262}^{*}$ & ${0.0095}$ & ${0.0111}^{*}$ & ${0.0342}^{*}$ & ${0.0539}^{*}$ & ${0.0179}^{*}$ & ${0.0230}^{*}$ \\
MMGCN$_{VLMo-ETE}$ & $\textbf{0.0319}$  & $\textbf{0.0522}$  &  $\textbf{0.0169}$ & $\textbf{0.0221}$ 
& $\textbf{0.0182}$ & $\textbf{0.0306}$ & $\textbf{0.0095}$ & $\textbf{0.0127}$ 
& $\textbf{0.0354}$ & $\textbf{0.0574}$ & $\textbf{0.0188}$ & $\textbf{0.0245}$ \\
\midrule
MMGCN$_{CLIP-PT}$ & ${0.0312}^{*}$  & ${0.0494}^{*}$  &  ${0.0165}^{*}$ & ${0.0216}^{*}$ & ${0.0163}^{*}$ & ${0.0253}^{*}$ & ${0.0090}^{*}$ & ${0.0101}^{*}$ & ${0.0376}^{*}$ & ${0.0606}^{*}$ & ${0.0198}^{*}$ & ${0.0257}^{*}$ \\
MMGCN$_{CLIP-FT}$ & ${0.0320}^{*}$  & ${0.0515}^{*}$  &  ${0.0176}^{*}$ & ${0.0229}^{*}$ & ${0.0173}^{*}$ & ${0.0263}^{*}$ & ${0.0099}^{*}$ & ${0.0114}^{*}$ & ${0.0379}$ & ${0.0598}^{*}$ & ${0.0201}$ & ${0.0257}$ \\
MMGCN$_{CLIP-ETE}$ & $\textbf{0.0376}$  & $\textbf{0.0592}$  &  $\textbf{0.0197}$ & $\textbf{0.0253}$ 
& $\textbf{0.0196}$ & $\textbf{0.0323}$ & $\textbf{0.0102}$ & $\textbf{0.0134}$ 
& $\textbf{0.0393}$ & $\textbf{0.0621}$ & $\textbf{0.0208}$ & $\textbf{0.0267}$ \\
\midrule
MMGCL$_{VLMo-PT}$ & ${0.0649}^{*}$  & ${0.0941}^{*}$ & \textbf{0.0371} & {0.0446} & $\textbf{0.0446}$ & ${0.0648}^{*}$ & $\textbf{0.0247}$ & $\textbf{0.0299}$ & ${0.0540}^{*}$ & ${0.0822}^{*}$ & ${0.0298}^{*}$ & ${0.0370}^{*}$ \\
MMGCL$_{VLMo-FT}$ & $\textbf{0.0667}$  & $\textbf{0.0980}$ & \textbf{0.0371} & \textbf{0.0452}$^{*}$ & ${0.0445}$ & $\textbf{0.0653}$ & ${0.0238}$ & ${0.0294}$ & $\textbf{0.0551}$ & $\textbf{0.0825}$ & $\textbf{0.0306}$ & $\textbf{0.0383}$ \\
MMGCL$_{VLMo-ETE}$ & ${0.0644}^{*}$  & ${0.0951}^{*}$ & {0.0356}$^{*}$ & {0.0435}$^{*}$
& ${0.0409}^{*}$ & ${0.0634}^{*}$ & ${0.0222}^{*}$ & ${0.0279}^{*}$ 
& ${0.0547}^{*}$ & ${0.0815}^{*}$ & ${0.0296}^{*}$ & ${0.0365}^{*}$ \\
\midrule
MMGCL$_{CLIP-PT}$ & ${0.0647}^{*}$  & ${0.0959}^{*}$ & {0.0367}$^{*}$ & {0.0447}$^{*}$ & ${0.0478}^{*}$ & ${0.0703}^{*}$ & ${0.0264}^{*}$ & ${0.0321}^{*}$ & ${0.0538}^{*}$ & ${0.0804}^{*}$ & ${0.0293}^{*}$ & ${0.0362}^{*}$ \\
MMGCL$_{CLIP-FT}$ & ${0.0677}^{*}$  & ${0.1007}^{*}$ & \textbf{0.0381} &{0.0466}$^{*}$ & ${0.0453}^{*}$ & ${0.0686}^{*}$ & ${0.0252}^{*}$ & ${0.0310}^{*}$ & ${0.0556}^{*}$ & ${0.0827}^{*}$ & ${0.0308}^{*}$ & ${0.0381}^{*}$ \\
MMGCL$_{CLIP-ETE}$ & $\textbf{0.0699}$  & $\textbf{0.1065}$ & {0.0374} & \textbf{0.0467} 
& $\textbf{0.0571}$ & $\textbf{0.0855}$ & $\textbf{0.0306}$ & $\textbf{0.0378}$ 
& $\textbf{0.0591}$ & $\textbf{0.0876}$ & $\textbf{0.0315}$ & $\textbf{0.0388}$ \\
\midrule
SLMRec$_{VLMo-PT}$ & {0.0615}$^{*}$  & {0.0919}$^{*}$ & ${0.0348}^{*}$ & ${0.0426}^{*}$ & {0.0450}$^{*}$ & {0.0653}$^{*}$ & {0.0252}$^{*}$ & {0.0304}$^{*}$ & {0.0561}$^{*}$ & {0.0823}$^{*}$ & {0.0302}$^{*}$ & {0.0374}$^{*}$ \\
SLMRec$_{VLMo-FT}$ & {0.0655}$^{*}$  & {0.0952}$^{*}$ & ${0.0370}^{*}$ & ${0.0447}^{*}$ & \textbf{0.0455} & {0.0655}$^{*}$ & \textbf{0.0259} & \textbf{0.0312} & \textbf{0.0588} & \textbf{0.0836} & \textbf{0.0314} & \textbf{0.0393}\\
SLMRec$_{VLMo-ETE}$ & \textbf{0.0678}  & \textbf{0.0994} & $\textbf{0.0379}$ & $\textbf{0.0461}$ 
& {0.0450} & \textbf{0.0681} & {0.0248}$^{*}$ & {0.0306}$^{*}$ 
& {0.0527}$^{*}$ & {0.0790}$^{*}$ & {0.0296}$^{*}$ & {0.0363}$^{*}$\\
\midrule
SLMRec$_{CLIP-PT}$ & {0.0630}$^{*}$  & {0.0909}$^{*}$ & ${0.0354}^{*}$ & ${0.0427}^{*}$ & {0.0451}$^{*}$ & {0.0671}$^{*}$ & {0.0249}$^{*}$ & {0.0304}$^{*}$ & {0.0568}$^{*}$ & {0.0821}$^{*}$ & {0.0310}$^{*}$ & {0.0381}$^{*}$ \\
SLMRec$_{CLIP-FT}$ & {0.0654}$^{*}$  & {0.0936}$^{*}$ & ${0.0369}^{*}$ & ${0.0441}^{*}$ & {0.0458}$^{*}$ & {0.0677}$^{*}$ & {0.0254}$^{*}$ & {0.0311}$^{*}$ & {0.0592}$^{*}$ & {0.0835}$^{*}$ & \textbf{0.0320} & \textbf{0.0395} \\
SLMRec$_{CLIP-ETE}$ & \textbf{0.0721}  & \textbf{0.1058} & $\textbf{0.0402}$ & $\textbf{0.0448}$ 
& \textbf{0.0569} & \textbf{0.0859} & \textbf{0.0309} & \textbf{0.0382} 
& \textbf{0.0597} & \textbf{0.0889} & {0.0314} & {0.0387}$^{*}$ \\
\midrule
LATTICE$_{VLMo-PT}$ & {0.0472}$^{*}$  & {0.0758}$^{*}$ & {0.0242}$^{*}$ & {0.0312}$^{*}$ & {0.0344}$^{*}$ & {0.0512}$^{*}$ & {0.0177}$^{*}$ & {0.0209}$^{*}$ & {0.0502}$^{*}$ & {0.0817}$^{*}$ & {0.0264}$^{*}$ & {0.0348}$^{*}$ \\
LATTICE$_{VLMo-FT}$ & {0.0456}$^{*}$  & {0.0749}$^{*}$ & {0.0240} & {0.0307} & {0.0320}$^{*}$ & {0.0501}$^{*}$ & {0.0161}$^{*}$ & {0.0207}$^{*}$  & {0.0503}$^{*}$ & {0.0808}$^{*}$ & {0.0254}$^{*}$ & {0.0346}$^{*}$ \\
LATTICE$_{VLMo-ETE}$ & \textbf{0.0676}  & \textbf{0.1012} & \textbf{0.0366} & \textbf{0.0451} & \textbf{0.0542} & \textbf{0.0788} & \textbf{0.0298} & \textbf{0.0361} & \textbf{0.0579} & \textbf{0.0889} & \textbf{0.0314} & \textbf{0.0393} \\
\midrule
LATTICE$_{CLIP-PT}$ & {0.0500}$^{*}$  & {0.0785}$^{*}$ & {0.0259}$^{*}$ & {0.0332}$^{*}$ & {0.0396}$^{*}$ & {0.0593}$^{*}$ & {0.0211}$^{*}$ & {0.0263}$^{*}$ & {0.0524}$^{*}$ & {0.0830}$^{*}$ & {0.0275}$^{*}$ & {0.0354}$^{*}$ \\
LATTICE$_{CLIP-FT}$ & {0.0493}$^{*}$  & {0.0781}$^{*}$ & {0.0257}$^{*}$ & {0.0332}$^{*}$ & {0.0386}$^{*}$ & {0.0578}$^{*}$ & {0.0197}$^{*}$ & {0.0248}$^{*}$ & {0.0520}$^{*}$ & {0.0812}$^{*}$ & {0.0273}$^{*}$ & {0.0348}$^{*}$ \\
LATTICE$_{CLIP-ETE}$ & \textbf{0.0679}  & \textbf{0.1015} & \textbf{0.0365} & \textbf{0.0451} & 
\textbf{0.0543} & \textbf{0.0789} & \textbf{0.0298} & \textbf{0.0361} & 
\textbf{0.0581} & \textbf{0.0892} & \textbf{0.0313} & \textbf{0.0393} \\
\bottomrule
\end{tabular}
\vspace{-4mm}
\end{adjustbox} 
\end{table*}

\zxrevise{In Section~\ref{sec:rq1}, we have successfully integrated the pre-trained LMM encoders into the multi-modal recommendation models and demonstrated their effectiveness. In this section, we investigate the impact of the fine-tuning and end-to-end training paradigms when incorporating CLIP and VLMo into the recommendation models.}
\zxrevise{Table~\ref{tab:ft} presents the performance changes observed in these recommendation models \zxrevise{over 240 cases} when using \ya{the} pre-trained and fine-tuned CLIP and VLMo as multi-modal feature extractors. 
\zxrevise{Each tested case corresponds to a specific model using either the PT or FT variants, as indicated in the relevant rows for each model in the table.}
\zxrevise{Within the table, PT/FT/ETE are the abbreviations for Pre-Training, Fine-Tuning and End-To-End, respectively.}
These experiments enable us to draw further conclusions about the effectiveness of \zxrevise{the} LMM encoders in the context of multi-modal recommendation.}
\zijunrevise{Table~\ref{tab:ft} \ya{shows} that in 70\% of the \zxrevise{cases} across all three datasets, \ya{the} fine-tuned LMM encoders \ya{exhibit} a significant improvement in recommendation performance compared to their pre-trained counterparts. This suggests that when \ya{the} LMM encoders are fine-tuned with \ya{the} recommendation datasets, they can be effectively transferred to the recommendation domain.}
\zxrevise{This fine-tuning 
enhances} visual and textual embeddings \zxrevise{by achieving a} deeper alignment between modalities. 
Consequently, \zijunrevise{these} enhanced, well-aligned item embeddings, result in more accurate and contextually relevant embeddings for multi-modal recommendation models.
However, we observe that there is no performance improvement for both MMGCL$_{CLIP-FT}$ and MMGCL$_{VLMo-FT}$ compared to their respective pre-trained {variants} on the Amazon Clothing dataset.
One potential reason for the {observed decrease in} performance could be that the fine-tuning process for this MMGCL model has led to overfitting the training data, causing a decrease in performance on the validation and test data. 
{This overfitting could occur if the MMGCL model becomes too specialised in capturing the patterns in the training data, resulting in a decreased ability to generalise to unseen validation and test data.}
In line with the observations from {Section~\ref{sec:rq1}}, we find that the LATTICE variants using {the} fine-tuned CLIP/VLMo {encoders} continue to underperform when compared to the ones using {the} pre-trained CLIP/VLMo on all three datasets.
This confirms our assumption that {this is caused by} the conceptual conflict between LATTICE, which aims to construct item-item graphs based on semantic similarities across different modalities, and the fine-tuned CLIP/VLMO. {Since the} fine-tuned CLIP/VLMO {encoders} can generate {more closely} aligned {multi-modal} features 
than {the} pre-trained ones, this conflict becomes more pronounced, potentially affecting the performance of LATTICE.

\zxrevise{Another setting not previously considered in the literature is the investigation 
of the impact of using an end-to-end training paradigm when incorporating CLIP and VLMo into the recommendation models.}
We perform experiments to \cmw{address} this gap {and to gauge} the impact of this training approach.
\craig{This important investigation allows us to} determine (1) which is the most effective training paradigm, \craig{and} (2) whether recommendation losses can further enhance multi-modal representation learning in the context of a multi-modal recommendation task.
Table~\ref{tab:ft} also presents a detailed comparison of the results between the two-stage training and end-to-end training approaches \zxrevise{across 240 cases}.
\zxrevise{These tested cases are differentiated by each model using FT and ETE variants, as indicated in the respective rows for each model within the table.}
{The end-to-end variants incorporating CLIP into the recommendation models exhibit enhanced performance in 83\% of cases across all three datasets compared to their respective fine-tuned ones, with 98\% of these cases showing significant improvement.}
In contrast, 
{end-to-end variants with VLMo integration do not show similar improvements and even {lead to} a decline in performance.}
This observation {suggests} that the end-to-end training paradigm facilitates a seamless integration of the CLIP encoder into the existing recommendation models, whereas it does not produce the same level of compatibility for VLMo.
The performance decline in {the} end-to-end VLMo integration might be attributed to its architecture, which comprises a unified multi-modal transformer \zijunrevise{with} modality-specific expert FFNs.
While fine-tuning VLMo updates both expert FFNs and the unified transformer for better feature alignment, end-to-end training may impede {the} effective gradient propagation through expert FFNs due to the recommendation loss (e.g. BRR loss).
This could result in the expert FFNs becoming less specialised in handling modality-specific features as they are updated alongside the multi-modal recommendation model, potentially leading to less effective embeddings.
Another interesting finding {from Table~\ref{tab:ft}} is that the LATTICE$_{VLMO/CLIP-ETE}$ {model} overcomes the inferior performance exhibited by both the pre-trained and fine-tuned versions {of the LATTICE model}, outperforming LATTICE$_{VLMO/CLIP-FT}$
{in \zxrevise{all} 
cases across the three datasets.}
This observation indicates that the end-to-end training paradigm effectively addresses the conceptual conflict between LATTICE and {the} {LMM} encoders by generating more effective multi-modal features adapted to \io{the} specific recommendation task, guided by the recommendation loss. 
This process results in more semantically informative item-item graphs for LATTICE.
This simultaneous learning process allows the multi-modal models to produce complementary representations, {thereby} minimising {the} potential conflicts between LATTICE's item-item graph learning and the multi-modal encoders' feature alignment. 
{Hence}, this integrated training strategy enables the LATTICE model to capitalise on the strengths of the multi-modal encoders, resulting in an improved performance.

\zijunrevise{\zxrevise{In answer to} RQ2, we \ya{found} that fine-tuning the LMM encoders improves performance in 70\% of the 240 
tested cases when comparing \ya{the} fine-tuned \zxrevise{LMM encoders} with \ya{the} pre-trained ones. However, some models exhibit anomalies due to inherent conceptual discrepancies between the fine-tuned encoders and their primary objectives. Moreover, all five models gain from an end-to-end training approach when integrating a dual-stream LMM encoder (i.e., CLIP), while the unified LMM encoder (i.e., VLMo) \ya{does not} show the same advantages.}
\zxrevise{This end-to-end training paradigm effectively \zxrevise{addresses} the conceptual conflict between the LMM encoders and \ya{a model such as} LATTICE.
Through this paradigm, we achieve a deeper feature alignment in multi-modal recommendation.}

\subsection{Modality Contribution Analysis (RQ3)}
\begin{table}[tb]
\centering
\caption{Comparative analysis of NDCG@20 scores for MMGCL and LATTICE models using different modality inputs across used datasets. V/T are the abbreviations for Visual/Textual, respectively. $^{*}$ indicates a significance difference using paired t-test with $p<0.05$.}
\label{tab:contr}
\begin{adjustbox}{width=\linewidth}
\begin{tabular}{lccccc}
\toprule
\multirow{1}{*}{\textbf{Dataset}} & \multicolumn{1}{c}{Amazon Sports} & \multicolumn{1}{c}{Amazon Clothing} & \multicolumn{1}{c}{Amazon Baby}\\ 
Methods & NDCG@20 & NDCG@20  & NDCG@20\\
\midrule
MMGCL$_{CLIP-ETE}$ (V\&T) & $\textbf{0.0467}$  & $\textbf{0.0378}$ & \textbf{0.0388} \\
MMGCL$_{CLIP-ETE}$ (V) & {0.0446}$^{*}$  & {0.0344}$^{*}$ & {0.0362}$^{*}$  \\
MMGCL$_{CLIP-ETE}$ (T) & {0.0455}$^{*}$  & {0.0367} & {0.0371}$^{*}$ \\
\midrule
LATTICE$_{CLIP-ETE}$ (V\&T) & \textbf{0.0451}  & \textbf{0.0361} & \textbf{0.0393} \\
LATTICE$_{CLIP-ETE}$ (V) & {0.0440}$^{*}$  & {0.0346}$^{*}$ & {0.0371}$^{*}$ \\
LATTICE$_{CLIP-ETE}$ (T) & {0.0449}  & {0.0355}$^{*}$ & {0.0379}$^{*}$ \\
\bottomrule
\end{tabular}
\end{adjustbox} 
\vspace{-4mm}
\end{table}


\zxrevise{As highlighted in Section~\ref{sec:intro} and detailed in Table 1 of the same section, \ya{the} existing models insufficiently investigate the interdependencies between modalities, hence exhibiting \ya{a} suboptimal performance when fusing multi-modal features. 
Hence, we conduct an analysis to investigate the contribution of each modality on the used Amazon datasets, particularly after incorporating \ya{the} LMM encoders into the existing models.
Table~\ref{tab:contr} presents the results of \ya{the} MMGCL and LATTICE models when fed with single or multiple types of modalities as input.
For conciseness, we report the \ya{results} of different types of modalities for only MMGCL and LATTICE here, as these are the two most effective baselines on the \zxrevise{used} datasets
(conclusions on the other models and metrics are similar).}
\zxrevise{Table~\ref{tab:contr} indicates that both MMGCL and LATTICE, when integrated with an LMM encoder and inputted jointly with visual and textual embeddings, outperform the same models using \zxrevise{a} single modality input.} \zijunrevise{This result, complementing the findings of Table~\ref{tab:analysis}, suggests that the LMM encoders \cmw{can} successfully exploit \zxrevise{the} deep alignment, confirming our \zxrevise{intuition} that the inclusion of additional modalities in a model should intrinsically augment the knowledge base, thereby enhancing \zxrevise{its} performance in the multi-modal recommendation task.}
\zxrevise{In response to RQ3, we conclude that \ya{the} LMM encoders can facilitate the \zijunrevise{deeper} alignment from diverse modalities, irrespective of the fusion methods employed, within the context of multi-modal recommendation systems.}



\section{Conclusions}
\looseness -1 In this study, we \yo{investigated} the effectiveness of \zxrevise{the} large multi-modal (LMM) encoders (\yo{namely}, CLIP and VLMo) for the multi-modal recommendation. 
\zxrevise{Specifically, 
we incorporated \ya{the} LMM encoders as a supplementary component into the multi-modal recommendation \zxrevise{models}
to enhance the user/item embeddings used by each recommendation model.
Our experimental results {show} that both the pre-trained and fine-tuned CLIP and VLMo encoders effectively extract and align visual and textual features from raw images and texts, \zxrevise{and significantly enhance} the performance 
\zxrevise{in four out of five}
state-of-the-art multi-modal recommendation models \zxrevise{we} tested. \ya{However, we also} \ya{observed} that \ya{for} certain model architectures {(e.g. LATTICE)}, this was not the case, due to conceptual conflicts between the fine-tuned LMM encoders and the {models'} inherent learning objectives. 
We also investigated different training paradigms \zxrevise{for the LMM encoders}. Our experiments \ya{showed} that end-to-end training is more suitable for the multi-modal recommendation task when incorporating \ya{a} dual-stream LMM encoder (i.e., CLIP) into the existing models, while \ya{a} unified LMM encoder (i.e., VLMo) does not exhibit the same benefits. 
Interestingly, our experiments showed that the end-to-end training addresses the conceptual conflict between the LMM encoders and LATTICE, highlighting the importance of adopting suitable training paradigms when incorporating \ya{the} LMM encoders into multi-modal recommendation systems.}
\zxrevise{Moreover, our in-depth analysis of the modality contribution \zxrevise{to the recommendation performance} highlights the capacity of \ya{the} LMM encoders to align different modalities, thereby enriching existing models with a spectrum of modalities as opposed to relying on a single modality.}
\zixuan{To conclude, our study \ya{showed} that the remarkable performance enhancements exhibited by \ya{the} LMM encoders in other tasks~\cite{bao2022vlmo,radford2021learning} are also observed in the recommendation domain, thereby warranting further investigation into their potential for the multi-modal recommendation task.}

\balance
\bibliographystyle{ACM-Reference-Format}
\bibliography{reference}
\end{document}